\documentclass[amsmath,aps,pra,reprint,superscriptaddress]{revtex4-2}

\pdfoutput=1
\usepackage[utf8]{inputenc}
\usepackage[T1]{fontenc}
\usepackage{lmodern}
\usepackage{microtype}

\usepackage{graphicx}
\usepackage{hyperref}
\usepackage{xcolor}

\definecolor{one}{HTML}{CC0000}

\hypersetup{
  colorlinks=true,
  citecolor=one,
  linkcolor=one,
  urlcolor=one
}

\newcommand*{\ev}[1]{\langle #1 \rangle}
\newcommand*{\bra}[1]{\langle #1 \rvert}
\newcommand*{\ket}[1]{\lvert #1 \rangle}

\begin{document}

\title{Mitigating depolarizing noise on quantum computers with
  noise-estimation circuits}

\author{Miroslav Urbanek}

\email[Corresponding author: ]{urbanek@lbl.gov}

\affiliation{Computational Research Division, Lawrence Berkeley
  National Laboratory, Berkeley, CA 94720, USA}

\author{Benjamin Nachman}

\affiliation{Physics Division, Lawrence Berkeley National Laboratory,
  Berkeley, CA 94720, USA}

\author{Vincent R. Pascuzzi}

\affiliation{Physics Division, Lawrence Berkeley National Laboratory,
  Berkeley, CA 94720, USA}

\author{Andre He}

\altaffiliation[Present address: ]{IBM Quantum, IBM T. J. Watson
  Research Center, Yorktown Heights, NY 10598 USA}

\affiliation{Physics Division, Lawrence Berkeley National Laboratory,
  Berkeley, CA 94720, USA}

\author{Christian W. Bauer}

\affiliation{Physics Division, Lawrence Berkeley National Laboratory,
  Berkeley, CA 94720, USA}

\author{Wibe A. de Jong}

\email[Corresponding author: ]{wadejong@lbl.gov}

\affiliation{Computational Research Division, Lawrence Berkeley
  National Laboratory, Berkeley, CA 94720, USA}

\begin{abstract}

A significant problem for current quantum computers is noise. While
there are many distinct noise channels, the depolarizing noise model
often appropriately describes average noise for large circuits
involving many qubits and gates. We present a method to mitigate the
depolarizing noise by first estimating its rate with a
noise-estimation circuit and then correcting the output of the target
circuit using the estimated rate. The method is experimentally
validated on the simulation of the Heisenberg model. We find that our
approach in combination with readout-error correction, randomized
compiling, and zero-noise extrapolation produces results close to
exact results even for circuits containing hundreds of $CNOT$ gates.

\end{abstract}

\maketitle

\section{Introduction}

Noisy intermediate-scale quantum (NISQ) computers~\cite{Preskill:2018}
are current and near-term quantum computers that are not
fault-tolerant. The presence of noise and errors limits their
utility. Even quantum algorithms designed for NISQ devices, for
example the variational quantum eigensolver~\cite{Peruzzo:2014}, are
hampered by imperfections of real devices. Error rates are still too
large to solve relevant scientific problems on existing quantum
computers. Consequently, there has been a lot of effort to reduce
noise and mitigate errors present on these devices.

An important class of errors are readout errors. They manifest
themselves as readouts of incorrect qubit values during a measurement,
e.g., reading one while the qubit is in the zero state and vice
versa. Readout errors can be successfully mitigated with readout-error
correction. Several method with varying degree of sophistication have
been developed~\cite{Kandala:2017, Maciejewski:2020, Bravyi:2020,
  Funcke:2020, Garmon:2020, Nachman:2020}.

Another large source of errors are gate errors. They can be classified
into coherent and incoherent errors. Coherent errors preserve state
purity. They are typically small miscalibrations in control
parameters. Coherent errors usually produce similar errors in
consecutive executions of a quantum circuit and lead to a systematic
bias in the output. Incoherent errors can be understood as either
coherent errors with randomly varying control parameters or as
processes that entangle the system with its environment. Incoherent
errors are easier to handle than coherent errors, because they can
often be modeled as depolarizing noise. A method for converting
coherent errors into incoherent errors is randomized
compiling~\cite{Wallman:2016, Cai:2019, Cai:2020}.

Another practical technique of error mitigation is zero-noise
extrapolation~\cite{Li:2017, Temme:2017, Dumitrescu:2018, Endo:2018,
  Kandala:2019, McArdle:2019, He:2020, Sun:2020, Fuchs:2020,
  Strikis:2020, Tomkins:2020}. A circuit is executed multiple times
with a varying degrees of noise and the measured output is
extrapolated to the zero-noise limit. Other mitigation methods have
been developed as well~\cite{BonetMonroig:2018, Schwenk:2018,
  Reiner:2018, Endo:2019, Otten:2019a, Otten:2019b, Otten:2019c,
  Murphy:2019, Sagastizabal:2019, Smart:2019, Song:2019, Zhang:2020,
  Roggero:2020}.

In this work, we introduce a new mitigation method. From a given
quantum circuit, which we call a target circuit, we construct a
circuit with a similar structure that we call an estimation
circuit. We execute the estimation circuit to measure the depolarizing
noise rate and then use the measured rate to correct the output of the
target circuit. We experimentally demonstrate that the combination of
readout-error correction, randomized compiling, mitigation with
estimation circuits, and zero-noise extrapolation produces results
that are very close to the exact results.

We first describe the method, introduce a simple class of estimation
circuits, and present our full mitigation approach. We then show
improvements obtained for our test case, which is a simulation of the
Heisenberg model.

\section{Methods}

A simple model of incoherent noise is the depolarizing noise model
given by~\cite{Nielsen:2010}
\begin{equation}
  \label{model}
  \epsilon(\rho) = (1-p) \rho + p \frac{I}{2^n},
\end{equation}
where $\epsilon$ denotes the noise channel, $\rho$ is the density
matrix, $p$ is the probabilistic error rate that depends on the device
and also on the circuit, and $n$ is the number of qubits. Notice that
if $\rho$ is initially a pure state, one can reconstruct the initial
state from the noisy density matrix. For $p > 0$, the initial pure
state is the state with the largest weight in $\epsilon
(\rho)$. Alternatively, if one knows $p$, the initial density matrix
$\rho$ can be reconstructed simply by calculating the inverse
$\epsilon^{-1} (\rho)$.

Observables are given by Hermitean operators acting on the system
Hilbert space. They can be decomposed into sums of strings of identity
and Pauli matrices,
\begin{equation}
  O = \sum_{i} c_i \prod_{j = 1}^n \sigma^{i, j},
\end{equation}
where $c_i$ are real coefficients and $\sigma^{i, j} \in \{I,
\sigma_x, \sigma_y, \sigma_z\}$ are identity or Pauli matrices acting
on qubit $j$. The trace of $S = \prod_{j = 1}^n \sigma^{i, j}$ is
either $\operatorname{tr} (S) = 2^n$ if $S$ is a product of identity
matrices or $\operatorname{tr} (S) = 0$ otherwise. The expectation
value of an observable $O$ for a state represented by a density matrix
$\rho$ is $\ev{O} = \operatorname{tr} (\rho O)$. The expectation value
of $O$ for a noisy density matrix~\eqref{model} is therefore given by
\begin{equation}
  \overline{\ev{O}} = \operatorname{tr} [\epsilon(\rho) O] = (1-p)
  \ev{O} + \frac{p}{2^n} \operatorname{tr} ({O}),
\end{equation}
where we denote the noisy expectation value by an overline. Notice
that $\overline{\ev{S}} = 1$ for strings $S$ consisting of identity
matrices only and $\overline{\ev{S}} = (1-p) \ev{S}$ otherwise.

We can therefore decompose any observable $O$ as $O = cI + O'$, where
$c$ is a constant, $I$ is the identity operator, and
$\operatorname{tr} (O') = 0$. Its expectation value is $\ev{O} = c +
\ev{O'}$. If we assume that the system decoherence is well described
by the depolarizing noise model and if we know $p$, we can correct a
noisy expectation value by calculating
\begin{equation}
  \label{correction}
  \ev{O} = \frac{\overline{\ev{O}} - c}{1-p} + c,
\end{equation}
where $\ev{O}$ is the corrected expectation value. We assume $c = 0$
in the following without loss of generality, because $c$ is just a
constant shift of the expectation value known in advance.

To correct the expectation value of any observable under the
depolarizing noise model, we have to estimate the value of $p$. We do
it by executing a circuit that is similar to our target circuit but
has a known output. We assume that the target circuit consists of
single-qubit and $CNOT$ gates only and that $CNOT$ gates are the
leading source of gate errors. Our approach to construct an estimation
circuit is to remove all single-qubit gates from the target circuit
and to keep only the $CNOT$ gates in it. Since the initial state on a
quantum computer is the zero state, ideal $CNOT$ gates do not
transform the initial state at all. The final state is again a zero
state on an ideal quantum computer. We can therefore estimate $1-p$ by
measuring the probability of obtaining the zero state with the
estimation circuit. The main assumption is the estimation and the
target circuit are affected by a similar $p$ because they have the
same structure.

It is not always necessary to remove all single-qubit gates. The
estimation circuit can be any circuit that has a known output
sensitive to noise and that has a similar structure as the target
circuit. It may be beneficial to preserve some single-qubit gates to
keep it similar to the target circuit. We add a layer of random
rotations as the first circuit layer and its inverse as the last
circuit layer to increase the robustness of the estimation.

Alternative recent approach uses near-Clifford circuits, which one can
simulate classically, to perform mitigation~\cite{Czarnik:2021}. The
main difficulty of this approach is that the output of random
near-Clifford circuits is similar to an output obtained with a
completely mixed density matrix. One therefore has to select a
particular subset of circuits that produce biased outputs. The authors
used machine learning to find appropriate near-Clifford circuits with
this property. Our method does not require any such selection. We
simply remove single-qubit gates to obtain a biased circuit that can
be simulated trivially.

We implemented our method in combination with readout-error
correction, randomized compiling, and zero-noise
extrapolation. Readout-error correction is performed using the
unfolding~\cite{Nachman:2020, Urbanek:2020} method.

\begin{figure}
  \centering
  \includegraphics{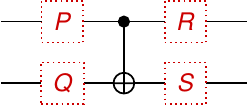}
  \caption{\label{compiling} Randomized compiling. Each $CNOT$ gate is
    dressed with the $P$, $Q$, $R$, and $S$ gates.}
\end{figure}

Coherent errors are dominant gate errors. They are not covered well by
the depolarizing noise model. Randomized compiling~\cite{Wallman:2016}
can convert coherent errors into incoherent errors. In particular, we
consider single-qubit gates being the easy gates and $CNOT$ gates
being the hard gates. We perform randomized compiling by inserting a
layer of randomizing single-qubit gates before and after each layer of
$CNOT$ gates as shown in Fig.~\ref{compiling}. The randomizing gates
are the identity and the Pauli gates. Each $CNOT$ gate is preceded and
succeeded by a pair of gates so that the overall action of the four
single-qubit gates and a $CNOT$ gate is exactly equal to a $CNOT$
gate. All possible gate choices are listed in Table~\ref{gates}. The
layer of single-qubit gates after a layer of $CNOT$ gates can be
composed with a layer of single-qubit gates before the next layer of
$CNOT$ gates. The circuit structure therefore consists of layers of
$CNOT$ gates interspersed with layers of single qubit gates. We use
randomized compiling for both the estimation and the target circuit.

\begin{table}
  \caption{\label{gates} Gate choices in randomized compiling. Each
    $P$, $Q$, $R$, and $S$ assignment produces a dressed gate equal to
    a $CNOT$ gate. An assignment is chosen independently and randomly
    and for each $CNOT$ gate in the circuit.}
  \begin{ruledtabular}
    \begin{tabular}{cccc@{\qquad}cccc@{\qquad}cccc@{\qquad}cccc}
      $P$ & $Q$ & $R$ & $S$ & $P$ & $Q$ & $R$ & $S$ & $P$ & $Q$ & $R$
      & $S$ & $P$ & $Q$ & $R$ & $S$ \\
      \colrule
      $I$ & $I$ & $I$ & $I$ & $Y$ & $I$ & $Y$ & $X$ & $X$ & $I$ & $X$
      & $X$ & $Z$ & $I$ & $Z$ & $I$ \\
      $I$ & $X$ & $I$ & $X$ & $Y$ & $X$ & $Y$ & $I$ & $X$ & $X$ & $X$
      & $I$ & $Z$ & $X$ & $Z$ & $X$ \\
      $I$ & $Y$ & $Z$ & $Y$ & $Y$ & $Y$ & $X$ & $Z$ & $X$ & $Y$ & $Y$
      & $Z$ & $Z$ & $Y$ & $I$ & $Y$ \\
      $I$ & $Z$ & $Z$ & $Z$ & $Y$ & $Z$ & $X$ & $Y$ & $X$ & $Z$ & $Y$
      & $Y$ & $Z$ & $Z$ & $I$ & $Z$ \\
    \end{tabular}
  \end{ruledtabular}
\end{table}

In the original formulation of zero-noise
extrapolation~\cite{Temme:2017, Kandala:2019}, the authors varied gate
duration assuming that gate errors increase with longer gates. They
ran experiments for several values of duration and extrapolated the
measured results to zero duration. A pulse-level control is required
to implement this method. This technique has been extended to systems
with gate-level control~\cite{Dumitrescu:2018}. The main assumption is
that $CNOT$ gates are the dominant source of errors. The authors
replaced each $CNOT$ gate with a sequence of three or five $CNOT$
gates, which are equivalent to a single $CNOT$ gate, executed their
circuits, and extrapolated to the zero-gate limit. This idea has been
further extended to replace only a subset of $CNOT$ gates with
sequences of $CNOT$ gates~\cite{He:2020}. Both methods assume a
certain dependence of errors on the number of $CNOT$ gates. In this
work, we execute three versions of each circuit, where each $CNOT$
gate is replaced by one, three, or five consecutive $CNOT$ gates, and
perform quadratic extrapolation to the limit corresponding to zero
$CNOT$ gates.

\section{Experiment}

Our test case is time evolution of the Heisenberg model. We consider a
quench of a one-dimensional XX chain of noninteracting spin-$1/2$
particles~\cite{Smith:2019}. Its Hamiltonian is given by
\begin{equation}
  H = -J \sum_{i = 1}^{n-1} \left( \sigma_x^j \sigma_x^{j+1} +
  \sigma_y^j \sigma_y^{j+1} \right),
\end{equation}
where $J > 0$ is a coupling constant, and $\sigma_x^i$ and
$\sigma_y^i$ are Pauli matrices acting on qubit $j$. The system is
initially prepared in a domain-wall configuration $\ket{\psi_0} =
\ket{\dots 111000 \dots}$ with qubits in the first and second half of
the chain in the one and zero state, respectively. We consider $J = 1$
and $\hbar = 1$ in the following for simplicity.

The propagator $\exp (-i H t)$ is approximated by its second-order
Trotter--Suzuki decomposition~\cite{Trotter:1959, Suzuki:1976} to
enable its implementation on a quantum computer. The approximated
propagator for one time step is given by
\begin{equation}
  \label{trotter}
  e^{-i H \Delta t} \approx e^{-i F \Delta t / 2} e^{-i G \Delta t}
  e^{-i F \Delta t / 2},
\end{equation}
where $F$ and $G$ contain terms in $H$ that act only on odd and even
bonds, respectively, and $\Delta t$ is a time step. Since all terms in
both $F$ and $G$ commute with each other, we can decompose the
exponentials in Eq.~\eqref{trotter} into products of two-qubit
exponentials of the form $\exp[-i (\sigma_x^j \sigma_x^{j+1} +
  \sigma_y^j \sigma_y^{j+1})\Delta t / d]$, where $d \in \{1,
2\}$. Each such exponential can be implemented by a circuit consisting
of two $CNOT$ gates and a number of single-qubit gates. One time step
is therefore implemented by three layers of two-qubit circuits acting
on odd, even, and odd bonds. Each two-qubit circuit is subsequently
decomposed into two $CNOT$ and multiple single-qubit gates. We measure
the time evolution of the local magnetization of the last spin in the
chain, $M_n(t) = \bra{\psi(t)} \sigma_z^n \ket{\psi(t)}$.

\begin{figure*}
  \centering
  \includegraphics{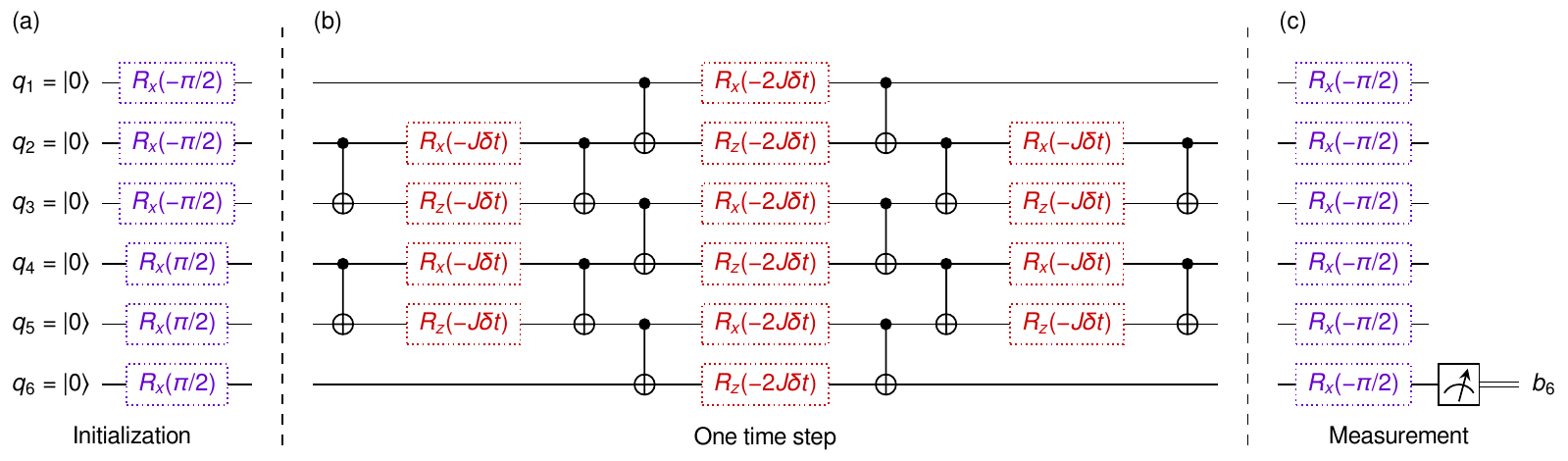}
  \caption{\label{circuit} Quantum circuit for the simulation of the
    XX chain. a) Preparation of the initial domain-wall state and
    basis transformation to a convenient basis. The dotted gates were
    replaced by random rotations in the estimation circuit. b) One
    step of the time evolution. Multiple steps are obtained by
    repeating this subcircuit. The dotted gates were removed in the
    estimation circuit. c) Basis transformation and measurement of the
    last qubit. The dotted gates were replaced by the inverses of
    random rotations from the initialization step in the estimation
    circuit.}
\end{figure*}

We implemented this model on the IBM Q Paris device using six qubits
Q23, Q24, Q25, Q22, Q19, and Q20 with 8192 shots for each circuit. The
circuit is shown in Fig.~\ref{circuit}. It contains 14 $CNOT$ gates
per time step. The longest circuit for 15 time steps contains 210
$CNOT$ gates.

We created the estimation circuit from the target circuit by removing
all its single-qubit gates. A layer of random single-qubit gates and a
layer of their inverses were added as the first and the last circuit
layer, respectively. Versions with one, three, and five $CNOT$ gates
per each $CNOT$ gate were created. We then constructed and executed
448 randomized instances of each circuit. Readout errors were
corrected using the unfolding technique as the first step in data
processing. All expectation values were averaged over the randomized
instances.

\begin{figure}
  \centering
  \includegraphics{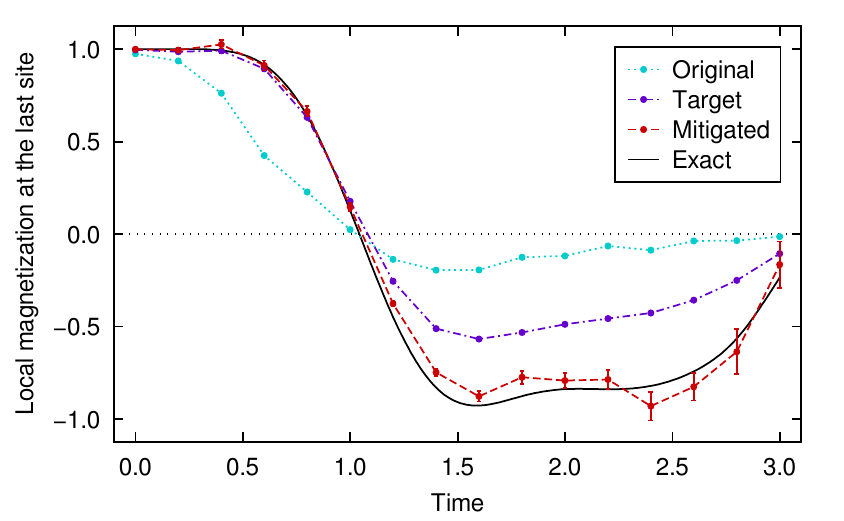}
  \caption{\label{xx} Comparison of the original and mitigated results
    for the time evolution of the local magnetization in the XX
    chain. The original results were obtained using the original
    circuits without any mitigation. There are 14 $CNOT$ gates per
    time step and the longest original circuit contains 210 $CNOT$
    gates. Target results use readout-error correction, randomized
    compiling, and zero-noise extrapolation. Mitigated results use
    readout-error correction, randomized compiling, mitigation with
    estimation circuits, and zero-noise extrapolation. Data for
    extrapolation were obtained with circuits where each $CNOT$ gate
    was replaced by one, three, or five $CNOT$ gates. Each circuit was
    executed with 448 random instances. Error bars represent the
    standard deviation of processed data. Exact solution takes the
    Trotter--Suzuki decomposition into account.}
\end{figure}

We estimated $1-p$ by measuring the $\ev{\sigma_z^6}$ expectation
value with estimation circuits. Ideally, $\ev{\sigma_z^6} = 1$, so the
depolarizing noise rate is given by $1-p = \ev{\sigma_z^6}$. The
mitigation was performed using Eq.~\eqref{correction}. We then applied
zero-noise extrapolation. Data points obtained with circuits with $n =
1$, $3$, and $5$ $CNOT$ gates were extrapolated to $n = 0$ using a
quadratic fit. The final results are shown in Fig.~\ref{xx}.

The improvement obtained using the zero-noise extrapolation on top of
the mitigation with estimation circuits can be understood as
follows. An expectation value measured with circuits with artificially
increased noise can be approximated by
\begin{equation}
  \label{noise}
  \overline{\ev{O}}_n = (1-p_n) \ev{O} \approx e^{-an} \ev{O},
\end{equation}
where $p_n$ is a noise rate, $\ev{O}$ is the exact expectation value,
$a$ is a constant that depends on the circuit, and $n$ is a noise
factor. In particular, $n = 1$, $3$, and $5$ for circuits with one,
three, and five $CNOT$ gates in place of each individual $CNOT$
gate. By estimating $1-p_n$ with the estimation circuit, we obtain a
value that is close to the true $1-p_n$ for the target circuit, but
not completely equal. However, the dependence on artificially
increased noise has the same form as in \eqref{noise},
\begin{equation}
  1-p_n \approx e^{-bn},
\end{equation}
where $b$ is a constant close to $a$. By performing mitigation with
estimation circuits, we obtain
\begin{equation}
  \ev{O}_n = \frac{\overline{\ev{O}}_n}{1-p_n} \approx e^{(b-a)n}
  \ev{O}.
\end{equation}
Since $|(b-a)n|$ is very small, we can approximate $\exp\left[ (b-a)n
  \right] \approx 1 + (b-a)n + (b-a)^2n^2/2$. The quadratic
extrapolation to $n = 0$ then works well, because this quadratic
function is a good approximation of the exponential function in the
neighborhood of zero. In contrast, zero-noise extrapolation using
expectation values obtained just with the target circuits does not
work well because the quadratic function does not approximate the
exponential well for large $an$. Another option may be to use an
exponential function instead of a quadratic function to directly fit
the data obtained with target circuits. We have found that this
approach is very sensitive to noise and leads to worse results than
the mitigation with estimation circuits. Fitting an exponential
function may work better with less noisy data.

\section{Conclusion}

We presented a method to mitigate errors and noise on quantum
computers that are described by the depolarizing noise model. The
method prescribes a construction of an estimation circuit to estimate
the noise rate that is then used to correct the output of a given
circuit. A crucial part of this approach is the randomized compiling
that ensures that gate errors can be modeled as incoherent
depolarizing noise. We demonstrated that the method works well,
especially in combination with readout-error correction and zero-noise
extrapolation, on a set of test circuits containing hundreds of $CNOT$
gates. The method is scalable to any number of qubits and gates given
that enough randomized samples are collected to achieve low
uncertainty.

\begin{acknowledgments}

This work was supported by the U.S. Department of Energy (DOE) under
Contract No.~DE-AC02-05CH11231, through the Office of Advanced
Scientific Computing Research Quantum Algorithms Team Program, and the
Office of High Energy Physics through the Quantum Information Science
Enabled Discovery program (Grant No.~KA2401032). This research used
resources of the Oak Ridge Leadership Computing Facility, which is a
DOE Office of Science User Facility supported under Contract
No.~DE-AC05-00OR22725.

\end{acknowledgments}

\appendix

\section{Zero-noise extrapolation}

\begin{figure}
  \centering
  \includegraphics{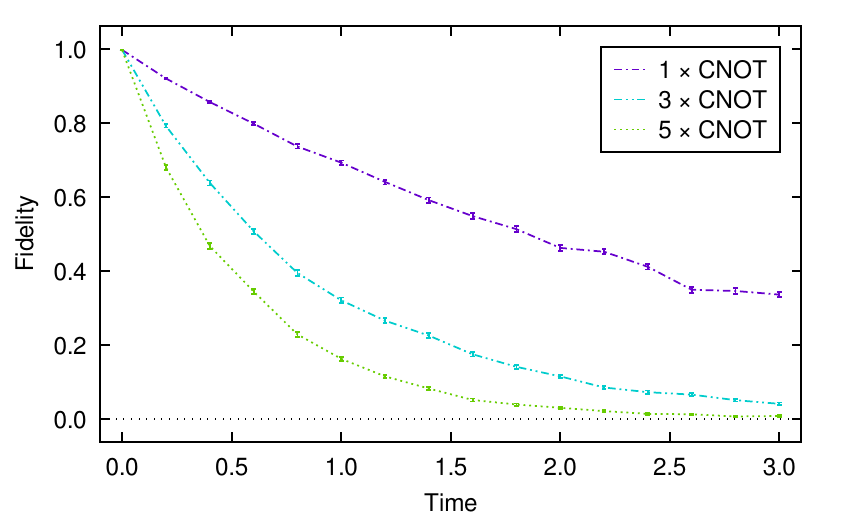}
  \caption{\label{estimation} Fidelity $1-p$ measured with the
    estimation circuits. Each point is a mean of values obtained by
    executing 448 randomized circuit instances. Error bars represent
    the standard error of the mean. Readout error correction was
    applied to measured data.}
\end{figure}

\begin{figure}
  \centering
  \includegraphics{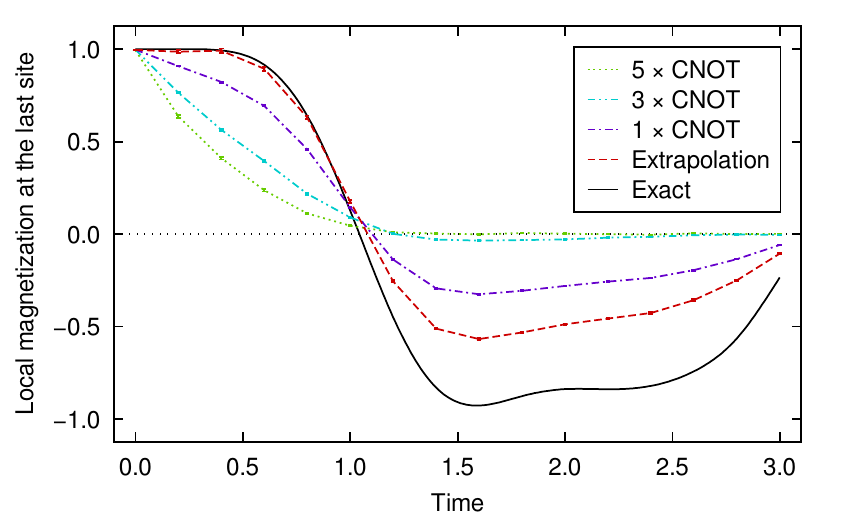}
  \caption{\label{target} Local magnetization measured with the target
    circuits. Each point is a mean of values obtained by executing 448
    randomized circuit instances. Error bars represent the standard
    error of the mean. Readout error correction was applied to
    measured data. Local magnetization was extrapolated to the
    zero-noise limit.}
\end{figure}

\begin{figure}
  \centering
  \includegraphics{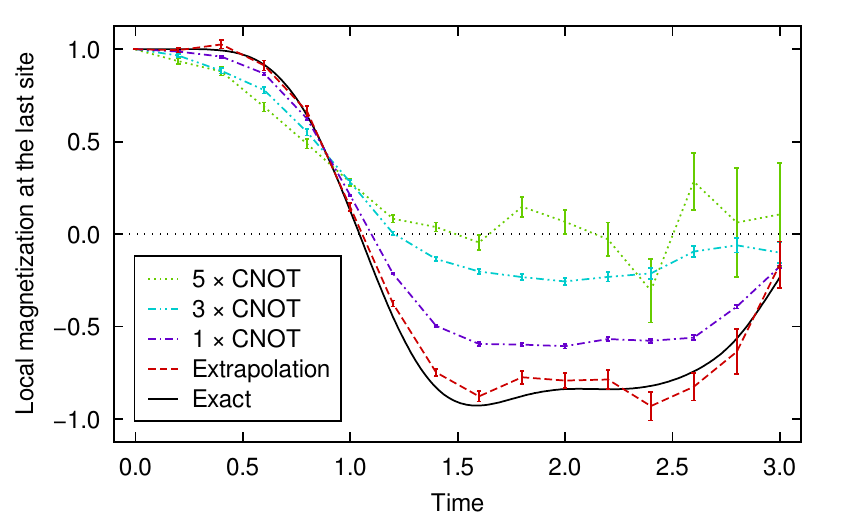}
  \caption{\label{mitigation} Mitigated local magnetization obtained
    from the the outputs of the estimation and target circuits. Error
    bars represent the standard deviation. Local magnetization was
    extrapolated to the zero-noise limit.}
\end{figure}

We replaced each $CNOT$ gate in our circuits with one, three, or five
consecutive $CNOT$ gates to artificially increase noise. They are
equivalent to a single $CNOT$ gate on a noiseless quantum
computer. Both the estimation and the target circuits were
modified. Figure~\ref{estimation} shows the effect of extra $CNOT$
gates on the output of the estimation circuits. Figure~\ref{target}
shows this effect on the output of the target circuits. We performed
mitigation using the outputs of the estimation and target circuits to
obtain mitigated local magnetization shown in Fig.~\ref{mitigation}.

\bibliographystyle{apsrev4-2}
\bibliography{main}

\clearpage

\end{document}